\documentstyle[12pt]{article}
\input{psfig}

\def\GeV{\,{\rm GeV}}

\def\keV{\,{\rm keV}}
\def\MeV{\,{\rm MeV}}
\def\sec{\,{\rm sec}}
\def\Gyr{\,{\rm Gyr}}

\def\Mpc{\,{\rm Mpc}}

\def\eV{{\,\rm eV}}

\def\cmm2{{\,\rm cm^{-2}}}
\def\cm2{{\,{\rm cm}^2}}
\def\cmm3{{\,{\rm cm}^{-3}}}
\def\gcmm3{{\,{\rm g\,cm^{-3}}}}
\def\kms{\,{\rm km\,s^{-1}}}

\def\la{\mathrel{\mathpalette\fun <}}

\def\fun#1#2{\lower3.6pt\vbox{\baselineskip0pt\lineskip.9pt
  \ialign{$\mathsurround=0pt#1\hfil##\hfil$\crcr#2\crcr\sim\crcr}}}

\begin{document}
\pagestyle{empty}
\begin{center}
\bigskip

%\rightline{FERMILAB--Pub--95/***-A}
%\rightline{astro-ph/9703174}
%\rightline{submitted to {\it }}

\vspace{.2in}
{\Large \bf COSMOLOGY:  FROM HUBBLE TO HST}
\bigskip

\vspace{.2in}
Michael S. Turner\\

\vspace{.2in}
{\it Departments of Physics and of Astronomy \& Astrophysics\\
Enrico Fermi Institute, The University of Chicago, Chicago, IL~~60637-1433}\\

\vspace{0.1in}
{\it NASA/Fermilab Astrophysics Center\\
Fermi National Accelerator Laboratory, Batavia, IL~~60510-0500}\\

\end{center}

\vspace{.3in}
%\centerline{\bf ABSTRACT}
%\bigskip
\begin{quote}
The Hubble constant sets the size and age of the Universe,
and, together with independent determinations of the age, provides a
consistency check of the standard cosmology.  The
Hubble constant also provides an important test of our most
attractive paradigm for extending the standard cosmology,
inflation and cold dark matter.
\end{quote}

\section{Introduction}

The value of the Hubble constant has changed by about a factor of ten since
Edwin Hubble's pioneering measurements.  The context in which we
view the Universe has changed just as profoundly.  Until 1964
cosmology was mostly concerned with cosmography; the spirit of
this period was perhaps best captured by Sandage, ``the quest
for two numbers ($H_0$ and $q_0$).''  The discovery of the Cosmic
Background Radiation led to the establishment of a physical foundation
for the expanding Universe -- the hot big-bang cosmology.
The 1970s saw this model become firmly established as the standard cosmology.
In the 1980s cosmologists began trying to extend the standard cosmology
by rooting it in fundamental physics.  Inflation is the first
step in this program.  Today, a host of cosmological observations
are testing inflation and its cold dark matter theory of structure formation;
here I focus on the role that the Hubble constant is playing in this enterprise.

\section{Foundations}

\begin{figure}[t]
\centerline{\psfig{figure=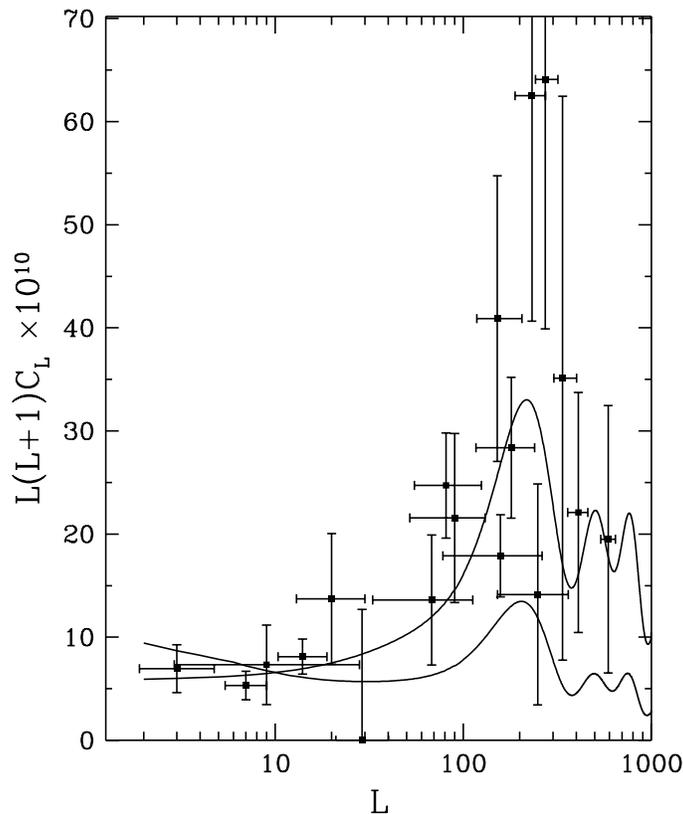,width=3.5in}}
\caption{Summary of CBR anisotropy measurements.
Plotted are the squares of the measured multipole amplitudes ($C_l = \langle
|a_{lm}|^2\rangle$) versus multipole number $l$.
The relative temperature difference on angular scale $\theta$ is
given roughly by $\protect\sqrt{l(l+1)C_l/2\pi}$
with $l\sim 200^\circ /\theta$. The theoretical curves are standard
CDM (upper curve) and CDM with $n=0.7$ and $h=0.5$ (lower curve).}
\end{figure}

The hot big-bang cosmology is a remarkable achievement.  It provides a
reliable accounting of the Universe from around $10^{-2}\sec$ until
the present, some $10\Gyr$ to $15\Gyr$ later.  It, together with the
standard model of particle physics and speculations about the unification
of the fundamental forces and particles, provides a firm foundation for
the sensible discussion of earlier times.

The standard cosmology rests on four observational pillars:
\begin{itemize}

\item The expansion of the Universe.  The redshifts and distances
of thousands of galaxies have been measured and are in accord with
Hubble's Law, $z = H_0 d$, a prediction of big-bang models for $z\ll 1$.

\item The Cosmic Background Radiation (CBR).  The
CBR is the most precise black body known -- deviations from
the Planck law are smaller than 0.03\% of the maximum intensity.  Its
temperature has been measured to four significant figures:
$T_0 = 2.728\pm 0.002\,$K \cite{firas}.  The only plausible origin
is the hot, dense plasma that existed in the Universe at times earlier
than $10^{13}\sec$ (epoch of last scattering and recombination).

\item Temperature fluctuations in the CBR.  Temperature differences
of order $30\mu$K between directions on the sky separated by angles
from less than one degree to ninety degrees have been measured by
more than ten different experiments \cite{white} (Fig.~1).  They establish the existence of
density inhomogeneities at the same level, $\delta \rho /\rho \sim
\delta T/T \sim 10^{-5}$, on length scales $\lambda \sim 100h^{-1}\Mpc
\,(\theta /{\rm deg})\sim 30h^{-1}\Mpc - 10^4h^{-1}\Mpc$.
Density perturbations of this amplitude, when amplified by the attractive
action of gravity over the age of the Universe, are sufficient to
explain the structure seen today.

\item Primeval abundance pattern of D, $^3$He, $^4$He and $^7$Li.
These light nuclei were produced a few seconds after the bang;
the predicted abundance pattern is consistent that seen in
primitive samples of the cosmos -- provided that the present
baryon density is between $1.5\times 10^{-31}\gcmm3$ and $4.5\times
10^{-31}\gcmm3$.  This corresponds to a fraction of critical
density $\Omega_Bh^2 = 0.008 - 0.024$ \cite{cst} (Fig.~2).
Nucleosynthesis is the earliest test of the hot big bang
and provides the best determination of the density of ordinary matter.

\end{itemize}

\begin{figure}[t]
\centerline{\psfig{figure=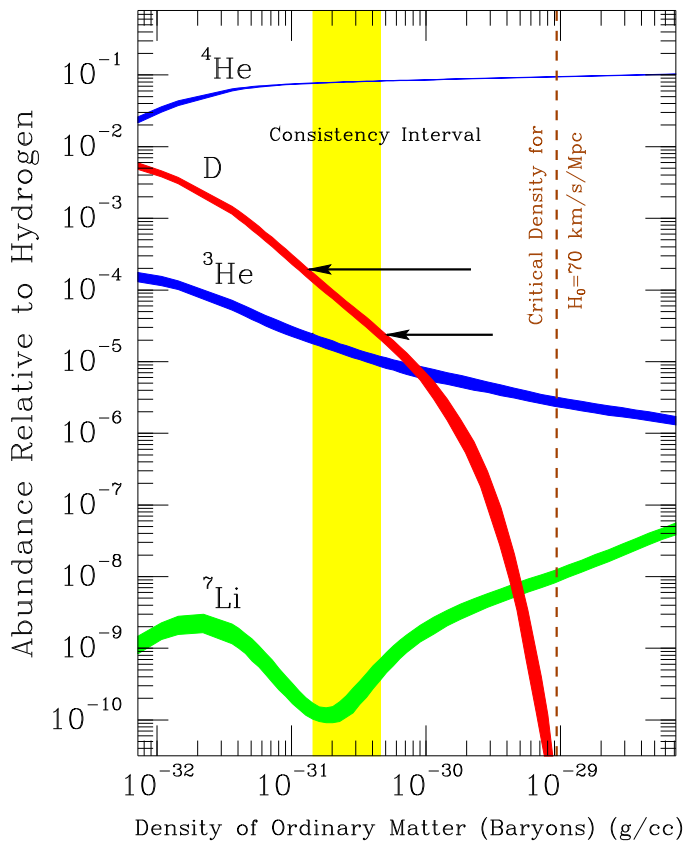,width=3.5in}}
\caption{Big-bang production of the light elements;
widths of the curves show the two-sigma theoretical uncertainty.
The primeval abundances of D, $^3$He, $^4$He and $^7$Li can be
explained if the baryon density is between $1.5\times
10^{-31}\gcmm3$ and $4.5\times 10^{-31}\gcmm3$ ($\Omega_Bh^2 =
0.008 - 0.024$).}
\end{figure}

The standard cosmology is successful in spite of our ignorance of the basic
geometry of the Universe -- age, size, and curvature -- which hinge
upon accurate measurements
of the Hubble constant and energy content of the Universe (fraction of
critical density in matter, radiation, vacuum energy, and so on).  The
expansion age, which is related to $H_0^{-1}$ and the energy content
of the Universe, is an important consistency check -- it should be
larger than the age of any object in the Universe.  The curvature radius
of the Universe is related to $H_0$ and $\Omega_0$:
$R_{\rm curv} = H_0^{-1}/\sqrt{|\Omega_0 -1|}$.

Note, the deceleration parameter is related to energy content of the
Universe, $q_0 = {1\over2} (\Omega_0 + 3\sum_i w_i\Omega_i )$, where
$\Omega_0$ is the total energy density divided the critical energy
density, $\Omega_i$ is the fraction of critical density in component
$i$ and $w_i$ is the ratio of the pressure contributed by component
$i$ to its energy density.  For a universe filled with nonrelativistic
matter, $q_0 = {1\over 2}\Omega_0$; for a universe with nonrelativistic
matter + vacuum energy (cosmological constant, $w_\Lambda = -1$),
$q_0 = {1\over 2}\Omega_0 -{3\over 2}\Omega_\Lambda$.

\section{Aspirations}

The hot big-bang model provides a firm physical basis for
the expanding Universe, but it leaves important questions unanswered.

\begin{itemize}

\item Quantity and composition of dark matter.  Most of the matter in
the Universe is dark and of unknown composition \cite{trimble}.
The peculiar velocities
of the Milky Way and other galaxies indicate that
$\Omega_{\rm Matter}$ is at least 0.3, perhaps as large as unity
\cite{pecvel}.  Luminous matter accounts for less mass density
that the lower limit to the baryon density from
nucleosynthesis ($\Omega_{\rm Lum}
\simeq 0.003h^{-1} < 0.008h^{-2} < \Omega_B$), and the upper limit to the
baryon density from nucleosynthesis is less
than 0.3 ($\Omega_B < 0.024h^{-2}<0.3$).  This defines the two dark-matter
problems central to cosmology (Fig.~3).  What is the nature of the dark
baryons? What is the nature of the nonbaryonic dark matter?

\item Formation of large-scale structure.  Gravitational
amplification of small primeval density inhomogeneities
provides the basic framework for understanding structure formation,
but important questions remain.  What is the origin of these
perturbations?  In detail, how did structure evolve?  The latter
is clearly tied to the dark-matter question.

\item Origin of matter-antimatter asymmetry.  During the earliest
moments ($t\la 10^{-6}\sec$), when temperatures
exceeded the rest-mass energy of nucleons, matter and antimatter existed
in almost equal amounts (thermal pair production made nucleons and
antinucleons as abundant as photons); today there is no antimatter and
relatively little matter
(one atom for every billion photons).  For this to be so, there must
have been a slight excess of matter over antimatter during the earliest
moments:  about one extra nucleon per billion nucleons and antinucleons,
for a net baryon number per photon of about $10^{-9}$.
What is the origin of this small baryon number?

\item Origin of smoothness and flatness.
Why in the large is the Universe so smooth (as evidenced by the CBR)?
The generic cosmological solutions to Einstein's equations are not smooth;
further, microphysical processes could not have smoothed things
out because the distance a light signal can travel at early times
covers only a small fraction of the Universe we can see.
Why was the Universe so flat in the beginning?  Had it not
been exceedingly flat, it would have long ago recollapsed
or gone into free expansion, resulting in a CBR temperature of
much less than 3\,K.

\item The beginning.  What launched the expansion?
What is the origin of the entropy (i.e., CBR)?  What was the big bang?
Is there a before the big bang?  Were there other bangs?
Are there more spatial dimensions to be discovered?

\end{itemize}

\begin{figure}[t]
\centerline{\psfig{figure=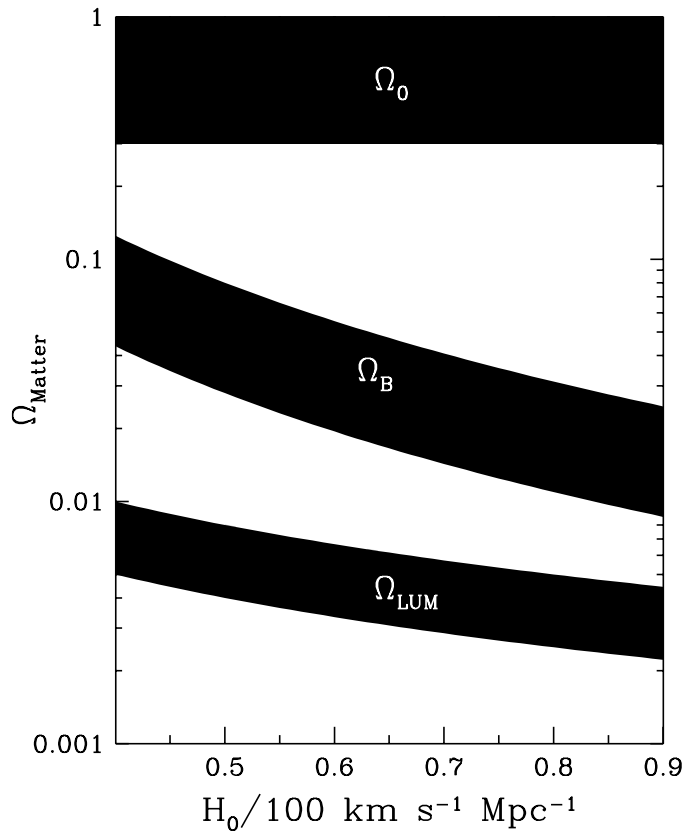,width=3.5in}}
\caption{Determinations of the matter density.
The lowest band is luminous matter, in the form of bright
stars and associated material; the middle band is the big-bang
nucleosynthesis determination of the density of baryons;
the upper region is the estimate of $\Omega_{\rm Matter}$
based upon the peculiar velocities of galaxies.  The
gaps between the bands illustrate the two dark matter problems:
most of the ordinary matter is dark and most of the matter is nonbaryonic.}
\end{figure}

This is an ambitious list of questions.  If physical explanations can be found,
we will have a more fundamental understanding of the Universe.
The study of the unification of the forces of Nature and the
application of these ideas to the early Universe has allowed
these questions to be addressed, and many of us believe that
answers will be found in the physics of the early Universe.  Over the past
fifteen years a number of important ideas have been put forth \cite{eu}
-- baryogenesis, topological defects (cosmic strings, monopoles,
textures, and domain walls), particle dark matter, baryogenesis, and
inflation.  I will
focus on inflation -- it is the most expansive, addresses almost
all the questions mentioned above, and is ripe for testing.

\section{Inflation and Cold Dark Matter}

Inflation \cite{inflation} holds that very early on (perhaps around $10^{-34}\sec$)
the Universe underwent a burst of exponential expansion driven by the
energy of a scalar field displaced from the minimum of its potential-energy curve.
(There are many candidates for the scalar field that drives inflation;
all involve new fields associated with physics beyond the standard
model of particle physics.)
During this growth spurt, the Universe expanded by a larger factor than
it has since.  Eventually the scalar field evolved to the minimum of its
potential and its energy was released into a thermal bath
of particles.  This entropy is still with us today:
the Cosmic Background Radiation.

The tremendous growth in size during inflation explains the large-scale
flatness and smoothness of the Universe:
After inflation, a very tiny patch of the pre-inflationary Universe,
which would necessarily
appear flat and smooth, becomes large enough to encompass all that
we see today and more.  Since spatial curvature and $\Omega_0$ are
related, inflation predicts a critical density Universe.\footnote{Recently,
it has been shown that inflation can accommodate
$\Omega_0 <1$, but at the expense of tuning precisely
the amount of inflation \cite{loomega}.}

The most stunning prediction of inflation is the linking of large-scale structure
in the Universe to quantum fluctuations on microscopic
scales \cite{scalar} ($\ll 10^{-16}\,$cm):  The wavelengths of quantum fluctuations
in the scalar field that drives inflation are stretched to astrophysical
size by the expansion that occurs during inflation.
The continual creation of quantum fluctuations and expansion
leads to fluctuations on all length scales; they develop into density
perturbations when the vacuum energy is converted
into radiation.  The spectrum is approximately scale invariant, that is,
fluctuations in the gravitational potential that are independent
of length scale.   The overall normalization of the spectrum is
dependent upon the shape of the scalar potential, and achieving
fluctuations of the correct size to produce the observed structure
in the Universe places an important constraint on it.

An inflationary model must incorporate two other pieces of early-Universe physics:
baryogenesis \cite{baryoreviews} and particle dark matter \cite{pdm}.
Since the massive entropy released
at the end of inflation exponentially dilutes any asymmetry that might have
existed between matter and antimatter, an explanation for the
matter -- antimatter asymmetry must be provided.
Baryogenesis is an attractive one.  It holds that particle interactions
that do not conserve baryon-number and do not respect $C$ and $CP$
(matter-antimatter) symmetry occurred out-of-thermal-equilibrium and
gave rise to the small excess of matter over antimatter needed to ensure
the existence of matter today.  Details of baryogenesis remain to
be worked out and tested -- did baryogenesis occur at modest temperatures
$T\sim 200\GeV$ and involve the baryon-number violation that exists
in the standard model or did it occur at much higher temperatures
and involve grand unification physics.

Particle dark matter is necessary since inflation predicts that
the Universe is at the critical density and baryons can contribute
at most 10\% of that.  While the standard model of particle physics
does not provide a particle dark matter candidate, many theories that
attempt to unify the forces and particles predict the existence
of new, long-lived particles whose abundance today is sufficient
to provide the critical mass density.  The three most promising
candidates are:  a neutrino of mass around $30\eV$; a neutralino
of mass between $10\GeV$ and $500\GeV$ \cite{neutralino}; and an axion
of mass between $10^{-6}\eV$ and $10^{-4}\eV$ \cite{axion}.

Inflation addresses essentially all the previously mentioned questions,
including the nature of the big bang itself.  As Linde \cite{linde} has emphasized,
if inflation occurred, it has occurred time and
time again (eternally to use Linde's words). What we refer to as the big bang is
simply the beginning of our inflationary bubble, one of an infinite number
that have been spawned and will continue to be spawned ad infinitum.
From the inflationary view, there is no need for a beginning.  (In that way,
inflation is similar to steady-state cosmology.)

There is no standard model of inflation, but there are a set of
robust predictions that allow inflation to be tested.

\begin{itemize}

\item Flat Universe.  Total energy density is equal to the critical density,
$\sum_i \Omega_i = 1$.  Among the components $i$ are baryons, slowly
moving elementary particles (cold dark matter), radiation (a very
minor component today, $\Omega_{\rm rad}\sim 10^{-4}$), and possibly
other particle relics or a cosmological constant.

\item Approximately scale-invariant spectrum of density perturbations.
More precisely, the Fourier components of the primeval density field
are drawn from a gaussian distribution with variance given by
power spectrum $P(k) \equiv \langle |\delta_k|^2 \rangle = Ak^n$ with
$n\approx 1$ ($n=1$ is exact scale invariance),
where $k=2\pi /\lambda$ is wavenumber and the
model-dependent constant $A$ sets the overall level of inhomogeneity
and is related to the form of the inflationary potential.

\item Approximately scale-invariant spectrum of gravitational waves.
Quantum fluctuations in the space-time metric give rise to
relic gravitational waves.
The overall amplitude of the spectrum depends upon the scalar potential
in a different way than the density perturbations.
These relic gravitational waves might
be detected directly by laser interferometers that
are being built (LIGO, VIRGO, and LISA) or by the
CBR anisotropies they produce \cite{lisa}.  If the spectra of both the matter
fluctuations and gravity waves can be determined, much could be learned
about the inflationary potential \cite{recon}.

\end{itemize}

The first two predictions lead to the cold dark matter (CDM) theory
of structure formation.\footnote{As a historical note the more conservative
approach of neutrino (hot) dark matter was tried first and
found to be wanting \protect\cite{nohdm}:  Since
neutrinos are light and move very fast they stream out of overdense
regions and into underdense regions, smoothing out density
inhomogeneities on small scales.  Structure forms from the top down:
superclusters fragmenting into galaxies -- which is inconsistent
with observations that indicate that superclusters are just forming today
and galaxies formed long ago.}
Within the cold dark matter theory, there are cosmological
quantities that must be specified in order to make precise
predictions \cite{dgt}.  They can be organized into two groups.
First are the cosmological parameters:  the
Hubble constant; the density of ordinary matter; the power-law index $n$ and
overall normalization constant $A$ that quantify the density perturbations;
and the level of gravitational radiation.\footnote{The level of gravitational
radiation is important because density perturbations are normalized by
CBR anisotropy and at present it is difficult to separate the contribution
of gravity waves to CBR anisotropy from that due to density
perturbations \cite{knox}.}
(A given model of inflation predicts $A$ and $n$ as well as the level
of gravitational radiation; however, there is no standard model of inflation.
Conversely, measurements of the above quantities can constrain -- and
even be used to reconstruct -- the scalar potential that
drives inflation \cite{recon}.)

The second group specifies the composition of invisible matter
in the Universe:  radiation, dark matter, and cosmological
constant.  Radiation refers to relativistic particles:  the photons in
the CBR, three massless neutrino species (assuming none of the neutrino
species has a mass), and possibly other undetected relativistic particles.
The level of radiation is crucial since it determines when the growth
of structure begins and thereby the shape of the power spectrum
of density perturbations today.  While the bulk of the dark matter
is CDM, there could be other particle relics;
for example, a neutrino species of mass $5\eV$, which
would account for about 20\% of the critical density.

The testing of cold dark matter began more than a decade ago with
a default set of parameters (``standard CDM'')
characterized by simple choices for both
the cosmological and the invisible matter parameters:
precisely scale-invariant density perturbations ($n=1$), $h=0.5$,
$\Omega_B =0.05$, $\Omega_{\rm CDM}=0.95$;
no radiation beyond photons and three massless neutrinos; no
dark matter beyond CDM; no gravitational waves; and zero cosmological constant.
The overall level of the matter inhomogeneity -- set by the
constant $A$ -- was fixed by comparing the predicted level of inhomogeneity
today with that seen in the distribution of bright galaxies.
Bright galaxies may or may not faithfully trace the distribution of mass.
In fact, there is some evidence that bright galaxies are
more clustered than mass, by a factor called the bias, $b \simeq 1 - 2$.
The distribution of galaxies today only fixes $A$ up to the bias factor $b$.

An important change occurred with the detection of CBR anisotropy
by COBE in 1992 \cite{dmr}.  The COBE measurement permitted a precise
determination of the amplitude of density perturbations on very large scales,
without regard to biasing.  And there was a surprise:
For standard CDM, the COBE normalization
predicts too much power on the scales of clusters and smaller \cite{jpo-ll}.

Figure 4 illustrates clearly that this problem simply reflects
a poor choice for the standard parameters.  It shows that there are
many COBE-normalized CDM models that are consistent with
measurements of the large-scale structure that exists today
(shape of the power spectrum of the galaxy distribution,
abundance of clusters, and early formation of structure in the
form of damped Lyman-$\alpha$ clouds; see Ref.~\cite{dgt}).  Organized
into families characterized by their invisible matter content they are:
CDM + cosmological constant ($\Lambda$CDM) \cite{lambda},
CDM + a small amount of hot dark matter ($\nu$CDM) \cite{nucdm},
CDM + additional relativistic particles ($\tau$CDM) \cite{taucdm},
and CDM with standard invisible matter content \cite{h30,stp}.

\begin{figure}[t]
\centerline{\psfig{figure=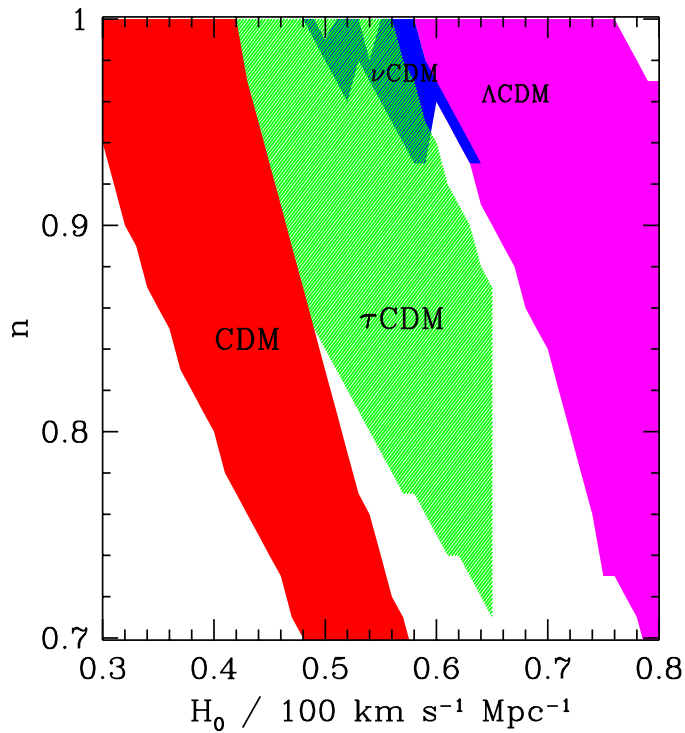,width=3.5in}}
\caption{Acceptable values of the cosmological
parameters $n$ and $h$ for CDM models with standard
invisible-matter content (CDM), with 20\% hot dark matter ($\nu$CDM),
with additional relativistic particles (the energy equivalent
of 12 massless neutrino species, denoted $\tau$CDM), and with a cosmological
constant that accounts for 60\% of the critical density ($\Lambda$CDM).
The $\tau$CDM models have been truncated at a Hubble constant
of $65\kms\Mpc^{-1}$ because a larger value would result in a
Universe that is younger than $10\Gyr$ (from Ref.~\protect\cite{dgt}).}
\end{figure}

\section{$H_0$ Tests Inflation and Cold Dark Matter}

A flood of cosmological observations -- from determinations
of the Hubble constant to measurements of CBR anisotropy -- are now
sharply testing inflation and cold dark matter.  Here I will focus
on the important role that $H_0$ plays.  It is two fold:
age-Hubble constant consistency and shape of the power spectrum
of inhomogeneity today (which depends upon $H_0$ as it determines the
value of the critical density and thereby the epoch of
matter-radiation equality).

The determinations of the ages of the oldest stars lie
between $12\Gyr$ and $17\Gyr$ \cite{yale,bolte}.
These estimates recent support from two other
independent methods -- the dating of the oldest white dwarfs based
upon how they cool and the dating of the radioactive elements, e.g.,
the isotope ratio of $^{235}$U/$^{238}$U \cite{truran}.  Taken
together, the case for an absolute minimum age of $10\Gyr$ appears ironclad.

On the other hand, measurements of the Hubble constant now favor
values between $60\kms\Mpc^{-1}$ and
$80\kms\Mpc^{-1}$, which for $\Omega_{\rm Matter} =1$ implies
an expansion age of $11\Gyr$ or less.  For a flat Universe with a cosmological
constant the expansion age is greater than ${2\over 3}H_0^{-1}$,
which lessens the age problem.  Within the uncertainties there is no
inconsistency, though there is tension, especially for models with
$\Omega_{\rm Matter} =1$ (Fig.~5).
Large-scale structure considerations ease the age problem,
as they favor an older Universe by virtue of a lower Hubble
constant or cosmological constant (Fig.~4).  Still, the Hubble constant
has great leverage.  Consider the following:

\begin{itemize}

\item $H_0 < 60\kms\Mpc^{-1}$.  CDM with standard invisible matter content
is viable.  However, the closer $H_0$ is to $60\kms\Mpc^{-1}$,
the more tilt (deviation of $n$
from unity) is required.  CBR anisotropy precludes $n$
less than 0.7; the next generation of satellite experiments, MAP
and COBRAS/SAMBA, should be able to determine $n$ to an accuracy of
a few percent.

\item $60\kms\Mpc^{-1}<H_0<65\kms\Mpc^{-1}$.  Only models with
nonstandard invisible-matter content are viable, e.g., $\nu$CDM and
$\tau$CDM.  $\nu$CDM has a smokin' gun signature:  around $5\eV$
worth of neutrino mass (in one or more species).  Particle-physics
models for producing extra relativistic particles ($\tau$CDM)
call for a massive ($1\keV - 10\MeV$), unstable tau neutrino.  There are a host
of laboratory experiments searching for evidence of neutrino mass.

\item $H_0 > 65\kms\Mpc^{-1}$.  Only $\Lambda$CDM is viable.  $\Lambda$CDM
too has a smokin' gun signature:  $q_0 = {1\over 2} - {3\over 2}\Omega_\Lambda
\approx -0.5$.  This should be tested soon by the two groups using
distant ($z\sim 0.3 - 0.7$) Type Ia supernovae to measure $q_0$.

\end{itemize}

Another test of inflation and CDM involves $H_0$, though less directly.
Because clusters of
galaxies are large objects it is expected that the cluster baryon
fraction, determined from x-ray measurements to be $(0.04 - 0.10)h^{-3/2}$
\cite{gasratio}, should closely reflect its universal value,
$\Omega_B/\Omega_{\rm Matter}$.  Using the nucleosynthesis value for
$\Omega_B$ fixes $\Omega_{\rm Matter}$ to be $(0.1-0.6)h^{-1/2}$.
Unless $H_0$ is very low, this determination of $\Omega_{\rm Matter}$
is only consistent with $\Lambda$CDM.  However, it should be remembered
that important assumptions must be made to infer the cluster baryon fraction --
that the hot intracluster gas is
unclumped and supported by thermal pressure alone -- if either is untrue
the actual baryon fraction would be smaller.

\begin{figure}[t]
\centerline{\psfig{figure=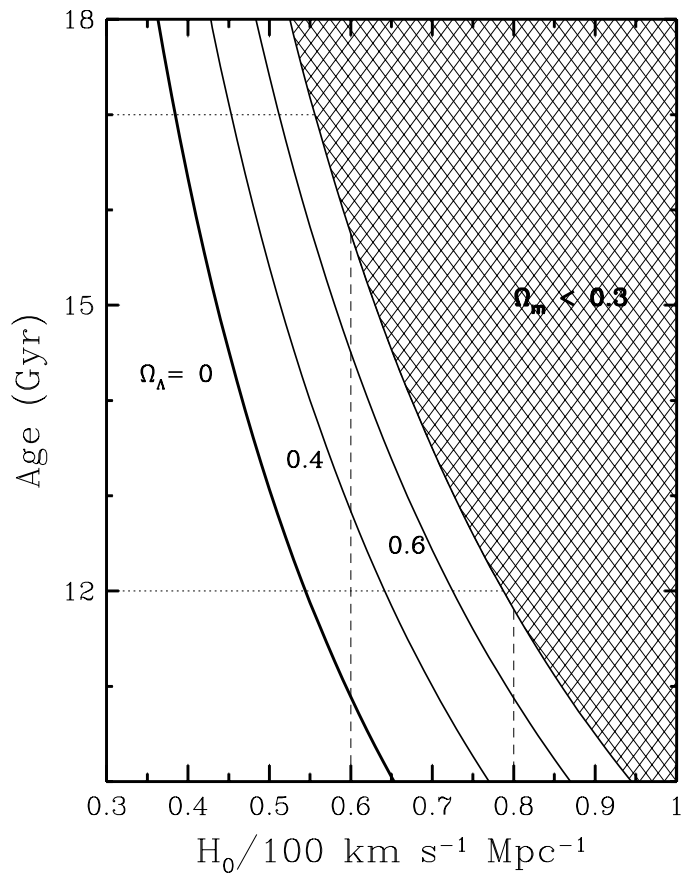,width=3.5in}}
\caption{The relationship between age and $H_0$ for flat-universe
models with $\Omega_{\rm Matter} = 1 - \Omega_\Lambda$.
The cross-hatched region is ruled out because
$\Omega_{\rm Matter} < 0.3$.  The broken lines indicate
the favored range for $H_0$ and for the age of the Universe.}
\end{figure}

\section{Concluding Remarks}

This is an exciting time in a cosmology.  We have a very successful
standard model, the hot big-bang cosmology, a bold and expansive
paradigm for extending it, inflation and cold dark matter, and the
observations that can test it are flooding in.  As I have
emphasized, the value of the Hubble constant has an
important role in this enterprise.  And of course,
the Hubble constant provides a consistency check
of the standard cosmology.

As we have heard at this meeting great progress is being made.
The Hubble Space Telescope is obtaining accurate
Cepheid distances to galaxies which can be used to calibrate
secondary indicators (e.g., supernovae of Types Ia and II, infrared
Tully-Fisher, and fundamental plane).  For the first time in
decades there is a consensus concerning the value of the Hubble constant:
$H_0 = 70 \pm 10 \kms\Mpc^{-1}$ (where $\pm 10\kms\Mpc^{-1}$ is indicative
of both the systematic and statistical errors).  You don't have to be
much of an optimistic to believe that a reliable determination
of the local Hubble constant to a precision of 10\% is within sight.

I believe that we will need to do better to really
test inflation.  A determination of the global Hubble constant
to a precision of 5\% will be needed and will require other
techniques.  The use of the aforementioned secondary indicators
can certainly determine $H_0$ out to 10,000 km/s, and perhaps to
30,000 km/s.  However, for a variety of CDM models (and probably
any model that reproduces the observed large-scale structure)
the one-sigma deviation of the local Hubble constant (within 10,000 km/s)
from its global value ranges from 4\% to 7\% \cite{shi,jpoturner}.
Moreover, there are uncertainties associated with the secondary
indicators that will be difficult to reduce below 5\%
(e.g., distance to LMC, Cepheid zero point, reddening corrections for
SN Ia, and so on).

The physically based methods -- time delays associated with gravitational
lenses, Sunyaev-Zel'dovich effect, and high-resolution mapping of
CBR anisotropy -- are well suited for this purpose.  First, they use
distant objects and thus probe the global $H_0$.
Next, the systematics are very different and probably less subject
to evolutionary and environmental effects.  Finally, as we have heard
at this meeting, their proponents believe that they are
capable of a 5\% determination.  In my opinion, CBR anisotropy
offers the most promise -- MAP and COBRAS/SAMBA have the potential to
make a one percent or better measurement of $H_0$.

Not having learned my lesson about speculating about the
value of the Hubble constant \cite{h30}, I reserve my final comments for
another try:  $53\kms\Mpc^{-1}$!  Let me assure the reader that
the explanation is more interesting than the value.
I take present measurements of the
Hubble constant to be $70\pm 10\pm 6\kms\Mpc^{-1}$
(where $\pm 6\kms\Mpc^{-1}$
reflects the one-sigma variance between the local and global values),
and further, use the following
prior information:  age of the Universe $t_0 = 15\pm 2\Gyr$, but necessarily
greater than $10\Gyr$; big-bang cosmology is correct, which, allowing
for a cosmological constant no larger than $\Omega_\Lambda = 0.7$,
implies $H_0t_0={1\over 2} - 1$.
The Bayesian probability distribution for $H_0$ is shown
in Fig.~6 -- it peaks around $60\kms\Mpc^{-1}$.  If I now include
a prior preference for the simplest CDM models
(those without nonstandard invisible matter) -- which requires a smaller value
of $H_0$, say $50\pm 10\kms\Mpc^{-1}$ (Fig.~4), and $H_0t_0 ={2\over 3}$ --
the probability distribution peaks around $53\kms\Mpc^{-1}$.
I note that physically based measurements of the Hubble constant, which
should reflect its global value, seem to be systematically
smaller, though they have larger errors, and thus give some
support to this value.

\begin{figure}[t]
\centerline{\psfig{figure=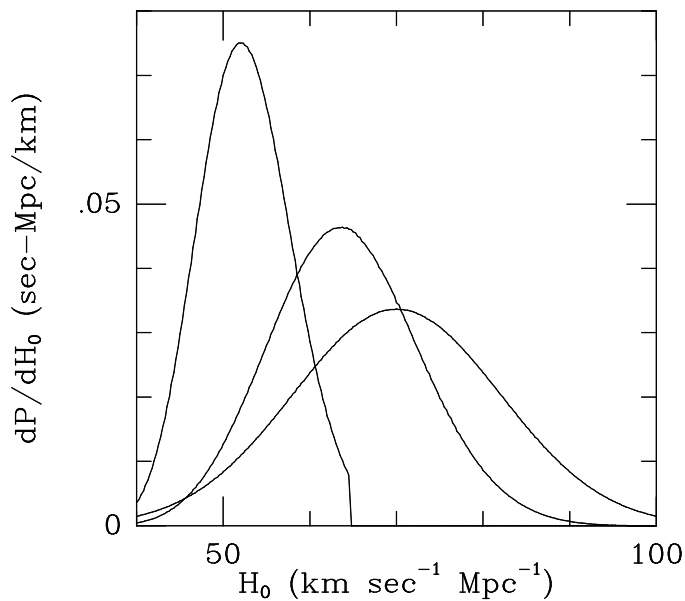,width=3.5in}}
\caption{Probability distributions for the global Hubble constant
based upon $H_0=70\pm 10\pm 6 \kms\Mpc^{-1}$ and different priors.
From right to left: no priors; priors on the age and correctness
of the big bang; priors on the age and correctness of the big bang and
and the simplest CDM models.}
\end{figure}

\paragraph{Acknowledgments.}  This work was supported by the DoE (at Chicago
and Fermilab) and by the NASA (at Fermilab by grant NAG 5-2788).

\end{document}